\documentclass{icrc2009}

\usepackage{graphicx}   % for including figures
\usepackage[caption=false]{caption}    % for captions
\usepackage[font=footnotesize]{subfig} % subfig.sty for a double column floating figure using two subfigures
\usepackage{fixltx2e}
\usepackage{url}

\newcommand{\shorttitle}[1]%
{\markboth{Proceedings of the 31\MakeLowercase{$^{st}$} ICRC, {\L}\'{o}d\'{z} 2009}{#1} }
\newcommand{\etal}{\MakeLowercase{\textit{et al. }}} % "et al."

\begin{document}
\title{A new upper limit on the redshift of PG 1553+113 from observations with the MAGIC Telescope}

\author{\IEEEauthorblockN{Elisa Prandini\IEEEauthorrefmark{1},
			  Daniela Dorner\IEEEauthorrefmark{2},
                          Nijil Mankuzhiyil\IEEEauthorrefmark{3},  \\
                          Mos\`e Mariotti\IEEEauthorrefmark{1} and
                          Daniel Mazin\IEEEauthorrefmark {4} for the MAGIC Collaboration}
                            \\
\IEEEauthorblockA{\IEEEauthorrefmark{1}Universit\`a di Padova and INFN Padova, I-35131 Padova, Italy}
\IEEEauthorblockA{\IEEEauthorrefmark{2}ETH Zurich, CH-8093 Switzerland}
\IEEEauthorblockA{\IEEEauthorrefmark{3}Universit\`a di Udine and INFN Trieste, I-33100 Udine, Italy}
\IEEEauthorblockA{\IEEEauthorrefmark{4}Institut de F\'isica d'Altes Energies, Edifici Cn., E-08193 Bellaterra (Barcelona, Spain)}
}

% please write the preseter's name and short title (3-4 words maximum)
%    which will appear at the header of the even pages.
\shorttitle{E.Prandini \etal PG~1553+113 redshift upper limit}
\maketitle

\begin{abstract}
Very high energy gamma ray emission from the active
 galactic nucleus PG~1553+113
 was observed during 2005 and 2006 by the MAGIC 
collaboration, for a total observation
 time of 18.8 hours. Here we
 present the results of follow up observations:
 more than 20 hours of good quality data 
collected by the MAGIC Telescope during the 2007 
and 2008 campaigns. The obtained spectra 
are compared and 
combined with previous measurements,
and corrected for absorption adopting different
 EBL models. Upper limits on the unknown 
source redshift are derived by assuming the 
absence of a break in the intrinsic spectrum,
or alternatively by constraining the hardness
 of the intrinsic source spectrum.

  \end{abstract}

 
\section{Introduction}
The BL~Lac object PG~1553+113, located at RA~15h55m43.0s, dec~+11d11m24s,
is one of the 27 Active Galactic Nuclei (AGNs) detected to date as Very
High Energy (VHE)
gamma ray emitters. AGNs are supermassive black holes surrounded by an accretion disk
and, in radio loud AGNs, bipolar jets of relativistic particles
  perpendicular to the disk plane. The spectral 
properties of the observed radiation are strictly related to the
viewing angle to the observer \cite{urri1995}.

AGNs whose jet points directly or at a small
angle to the observer belong to the class of blazars. 
BL~Lac objects, like PG 1553+113,
 are blazars showing very weak emission lines. 
The observed spectrum from a BL~Lac is totally dominated by the jet, 
since the power of the radiation emitted by the jet 
is enhanced by relativistic beaming effects. Typically, radiation emitted
from these objects covers the entire electromagnetic spectrum, from
radio wave to gamma-ray frequencies. The spectrum is composed of two bumps: one at
low energy and a second at high energy, peaking in the GeV range.
The first component is identified as electron synchrotron radiation, whilst the
nature of the second component is still under debate.
The most popular models of GeV component, the so-called leptonic models, invoke inverse
Compton scattering of ambient photons on electrons. 
Alternatively, the high energy photons observed in BL~Lac spectra could be of 
hadronic origin through the emission of secondary electrons.

\subsection{The redshift of PG~1553+113}

The distance of PG~1553+113 is still unknown.
Despite several observing campaigns with optical
instruments, no emission or absorption lines have been 
detected. Moreover, the observed jet emission from this source
is so bright that it prevents optical observation
of the host galaxy, which was recently attempted using the Hubble Space Telescope
 and the ESO Very Large Telescope.
These measurements resulted in a lower limit 
on the redshift of $z > 0.78$ and  $z > 0.09$ respectively 
\cite{sbarufatti2005,sbarufatti2006}. 
Under the assumption that the luminosity of the host galaxy of BL Lac 
objects can be considered constant, 
a lower limit of $z > 0.25$ was recently reported \cite{treves2007}.

The distance of a gamma-ray emitting extragalactic object is
of crucial importance for VHE observations. The presence of 
a diffuse optical/near-infrared background in intergalactic
space, the so-called Extragalactic Background Light (EBL), 
causes a partial/total absorption of the VHE photons coming
from distant sources. The most distant object with known redshift
 detected so far in VHE is the FSRQ 3C279, located at redshift $z=0.54$.

The detection of $\gamma$ emission above 100 GeV from the source
of unknown redshift PG~1553+113 has been reported 
using the H.E.S.S. and MAGIC telescopes \cite{1553HESS,1553MAGIC}.
This has allowed the development of new
 methods to determine the source distance,
based on the source spectral features \cite{aharonian2006,mazin2007}.
An upper limit on the redshift can be inferred by requiring that the
VHE component of the EBL corrected (i.e.\ deabsorbed) spectrum 
satisfies particular physical conditions.
A combination of MAGIC and H.E.S.S. spectra has been used to
set upper limits of $z<0.69$ and $z<0.42$, depending
on the intrinsic spectrum assumed \cite{mazin2007}.

In this paper, we set new upper limits
on the redshift of PG~1553+113 from the observed VHE spectrum.
We combine previously published 2005-2006 MAGIC data with
new MAGIC data taken in 2007 and 2008.
We utilise the lower limit EBL model from \cite{kneiske2004},
 which is at the level of direct lower limit set by
galaxy counts (see Fig.~\ref{ebl_distribution})
 and for comparison the recent model \cite{franceschini2008}, 
based on real data.
 
\section{Observations and data analysis \label{sec:obs}}
 \subsection{The MAGIC Telescope}

MAGIC \cite{baixeras04} is a new generation
Imaging Atmospheric Cherenkov Telescope located on La Palma,
Canary Islands, Spain ($28.3^{\circ}$N, $17.8^{\circ}$W, 2240\,m~asl).
Due to its low energy trigger threshold of 60\,GeV, 
MAGIC is well suited to multiwavelength observations together with instruments
operating in the GeV range. The parabolically-shaped reflector,
with a total mirror area of 236\,m$^{2}$ allows MAGIC to sample a
part of the Cherenkov light pool and focus it onto a multi-pixel
camera, composed of 576 ultra-sensitive photomultipliers.
The total field of view of the camera is $3.5^{\circ}$ and the collection area is
of the order of $10^{5}$\,m$^{2}$ at 200\,GeV for a source close to zenith.
The incident light pulses are converted into optical signals,
transmitted via optical fibres and digitised by 2\,GHz flash
ADCs \cite{goebel2007}.

 \subsection{Data analysis}

PG~1553+113 was observed with the MAGIC Telescope for 
nearly 19 hours in 2005 and 2006 \cite{1553MAGIC}. It was also
the subject of a multi-wavelength campaign carried out in July 2006
with optical, X-ray and TeV $\gamma$-ray telescopes \cite{albert2009}. 
Follow-up observations with the MAGIC telescope were carried out for 14 hours
in March-April 2007 and for nearly 26 hours in March-May 2008, 
parts of which were taken simultaneously with other 
instruments \cite{nijil2009}.
Unfortunately the 2008 observations were severely affected 
by bad weather (including \textit{calima}, Saharan sand-dust in the atmosphere)
 that greatly reduced the final dataset and resulted in an
increased energy threshold. Data from both periods were taken in the 
false-source tracking (wobble) mode \cite{fomin1994},
 in which the telescope was alternated every 20 minutes between
two sky positions at $0.4^\circ$ offset from the source.
The zenith angle of the 2007 observations varied
 from $17^\circ$ to $30^\circ$, while in 2008
it extended to $36^\circ$.

The analysis was performed using the standard MAGIC analysis
software \cite{bretz2005}. 
In the early stages of the analysis, an absolute 
calibration with muons
 and an absolute mispointing correction 
were  performed. The arrival time information for
pulses in neighbouring pixels was later used to suppress the
contribution from the Night Sky Background (NSB)
 in the shower images \cite{aliu2009}.
After these steps, the camera images were parameterised via the so-called
Hillas image parameters \cite{hillas1985}. Two additional parameters,
namely the time gradient along the main shower axis and
the time spread of the shower pixels, were computed \cite{albert2008a}.

Severe quality cuts based on event rate after NSB suppression were 
applied to the sample. 20.2 hours of good quality
 data remained after these cuts, of which 11.5 hours were taken in 2007 and 
8.7 hours in 2008. An additional cut removed the events with total charge less
than 80 photoelectrons (phe) in 2007 to ensure a 
better background rejection.
A harder cut, at 200 phe, had to be applied to the 2008 dataset
due to the poor observing conditions. For the 
successive steps of the analysis, Monte-Carlo simulations
of $\gamma$-like events were used. Hadronic background suppression was achieved
using the Random Forest (RF) method \cite{breiman2001},
in which each event is assigned an additional parameter, the hadronness,
which is related to the probability that the event is not $\gamma$-like.
The RF method was also used in the energy estimation.
The energy threshold was estimated to be 
80\,GeV in 2007 and 150\,GeV in 2008. 
Due to a change in the telescope performance,
 the optical point-spread function (PSF) of the two 
periods differs: the $\sigma$ of the PSF was measured to be 13\,mm and 
10\,mm in 2007 and 2008 respectively. To take this into account, data were 
analysed separately.
Effects on the spectrum determination introduced by the limited energy resolution 
were corrected by ``unfolding'' the final spectra.

\subsection{Results}

\begin{figure}[b!]
  \centering
  \includegraphics[width=3.0in]{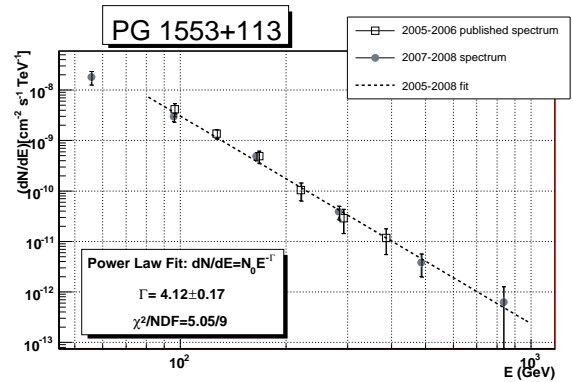}
  \caption{Combined differential measured energy spectrum of PG~1553+113.}
  \label{2005_2008_spectra_all}
 \end{figure}
 
\begin{figure*}[t!]
  \centering
  \includegraphics[width=5in]{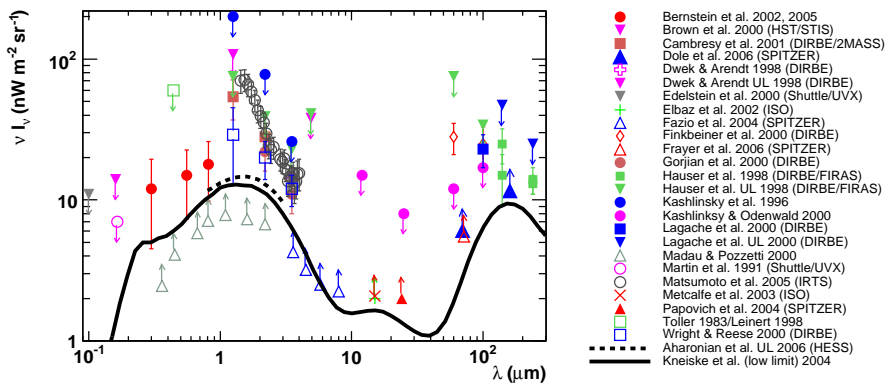}
  \caption{Energy density of the Extragalactic Background Light.
    Direct measurements, galaxy counts, low and upper limits are shown
    by different symbols. The dashed black line represents the upper limit
    set by H.E.S.S. \cite{aharonian2006}. The black solid curve
    is the minimum EBL spectrum at z=0 from the model adopted in this work.}
  \label{ebl_distribution}
 \end{figure*}

The 20.2 hours of observations of PG~1553+113 
carried out between 2007 and 2008 resulted in a signal
of $12\sigma$ significance,
in the energy range between 150 and 600\,GeV. 
The differential spectra measured by MAGIC in 2007 and 2008
were fitted 
with a power law function of the form
 \begin{equation}
 \frac{dF}{dE}= f_0 * {\left( \frac{E}{200\,{\rm GeV}}\right)}^{-\alpha}
 \label{simpeq}
\end{equation}
where $f_0$ is the flux at 200\,GeV and $\alpha$ is the
power law index. 
The resulting indices are listed in Table~\ref{table_spectra}, along with
the integral fluxes above 200\,GeV, as estimated from the fit.
The systematic uncertainty is estimated to be $35\%$ in the
flux level and 0.2 in the power index \cite{MAGICcrab}. 

  \begin{table}[h!]
  \caption{PG~1553+113 MEASURED SPECTRUM}
  \label{table_spectra}
  \centering
  \begin{tabular}{c|c|c}
  \hline
   Year   & F(E $>$ 200 GeV) & photon index  \\
   & $(10^{-11} cm^{-2} s^{-1})$& \\
   \hline 
    2007 & $ 0.69 \pm 0.13  $& $4.20 \pm 0.28$\\
 \hline 
    2008 & $ 1.76 \pm 0.24   $& $3.95 \pm 0.32$\\
 \hline 
    2007+2008 & $ 1.12 \pm 0.19   $& $4.03 \pm 0.23$\\
 \hline 
    2005+2006\footnotemark &  $ 1.0 \pm 0.4 $ & $4.21 \pm 0.25$ \\
 \hline 
    2005-2008 &  $ 1.13\pm 0.14  $ & $4.12 \pm 0.17$ \\
  \hline 
   \end{tabular}
  \end{table}
  \footnotetext{published data \cite{1553MAGIC}}

The absolute flux above 200\,GeV observed in 2008 spectrum
is a factor 2.5 larger compared to the one measured in 2007.
The differential spectrum seems
harder in 2008 than in 2007, though consistent within
the errors.
The absolute flux above 200\,GeV from the combined dataset
is in good agreement with previous measurements carried out by the 
MAGIC Telescope in 2005 and 2006 \cite{1553MAGIC}. Hence, all data
were combined in a single dataset, in order to increase statistics.
The overall spectrum, drawn in Fig.\ref{2005_2008_spectra_all} 
can be well described by a simple power law fit of steep index 
$4.12 \pm 0.17$.

\section{Absorption of VHE gamma-rays}
The VHE spectrum emitted by distant sources 
is strongly affected  by the interaction 
with the EBL \cite{hauser2001}.
EBL is composed of stellar light emitted and partially 
reprocessed by dust throughout the entire history of 
galaxy evolution. The expected spectrum of this light 
is composed by two bumps at near-infrared and far-infrared
wavelengths \cite{hauser2001}. Direct measurement of the EBL 
has proven to be a difficult task, primarily
due to the zodiacal light that forms a bright
foreground which is difficult to suppress. 
In Fig. \ref{ebl_distribution}, from \cite{mazin2007},
a collection of recent experimental data is shown.

VHE $\gamma$-rays from distant sources interact
 with the low-energy photons of EBL through electron-positron
pair production. As a result, VHE spectra from AGNs 
are  exponentially attenuated by a factor $\tau_{\gamma\gamma}(E,z)$,
where $\tau$ is the optical depth and is a function of 
both energy and redshift of the source.  
The EBL wavelength range for this 
absorption of VHE $\gamma$-rays extends from the UV to the far-infrared.
We adopt the optical depth values 
from \cite{kneiske2004} and from the 
recent model \cite{franceschini2008} 
to correct the observed spectra from PG~1553+113
for EBL absorption and derive a new upper limit on the redshift.
Both models take EBL evolution into account. 

\section{Upper limits on the redshift of PG~1553+113}
For the determination of the upper limit on the 
redshift of the source PG~1553+113, we decided to 
use two different approaches. The first is to require a
\emph{minimum value for the power law index}
 of the deabsorbed spectrum. According to the "standard 
scenario" of particle acceleration, the minimum allowed
index in AGN spectra is $\Gamma_{int}^{st}=1.5$ 
\cite{aharonian2006}. Several theoretical possibilities
have been proposed to create harder spectra. 
As an "extreme case" we also consider
the value $\Gamma_{int}^{ex}=2/3$, discussed in \cite{katar2006}.
The second method, recently adopted in \cite{mazin2007},
is based on the hypothesis that there is no
break in the intrinsic VHE source spectrum.
\begin{figure*}[!t]
   \centerline{\subfloat[Constraints on the reshift of PG~1553+113 with 
       the likelihood ratio test. See text for details.]
     {\includegraphics[width=2.5in]{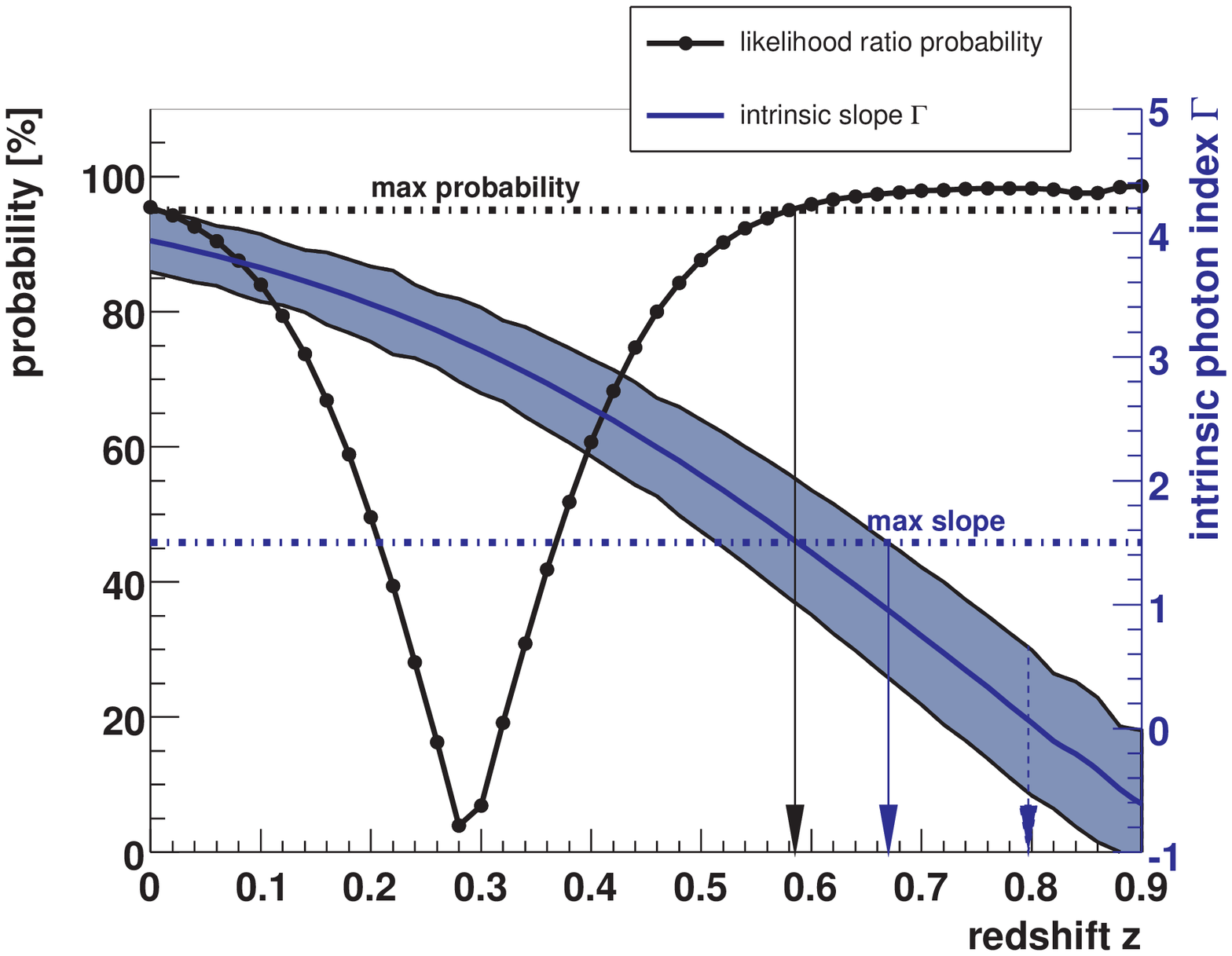} \label{scan}}
     \hfil
     \subfloat[Intrinsic spectrum of PG~1553+113 at redshift 0.58, using minimum EBL model. See text for details.]{\includegraphics[width=3.0in]{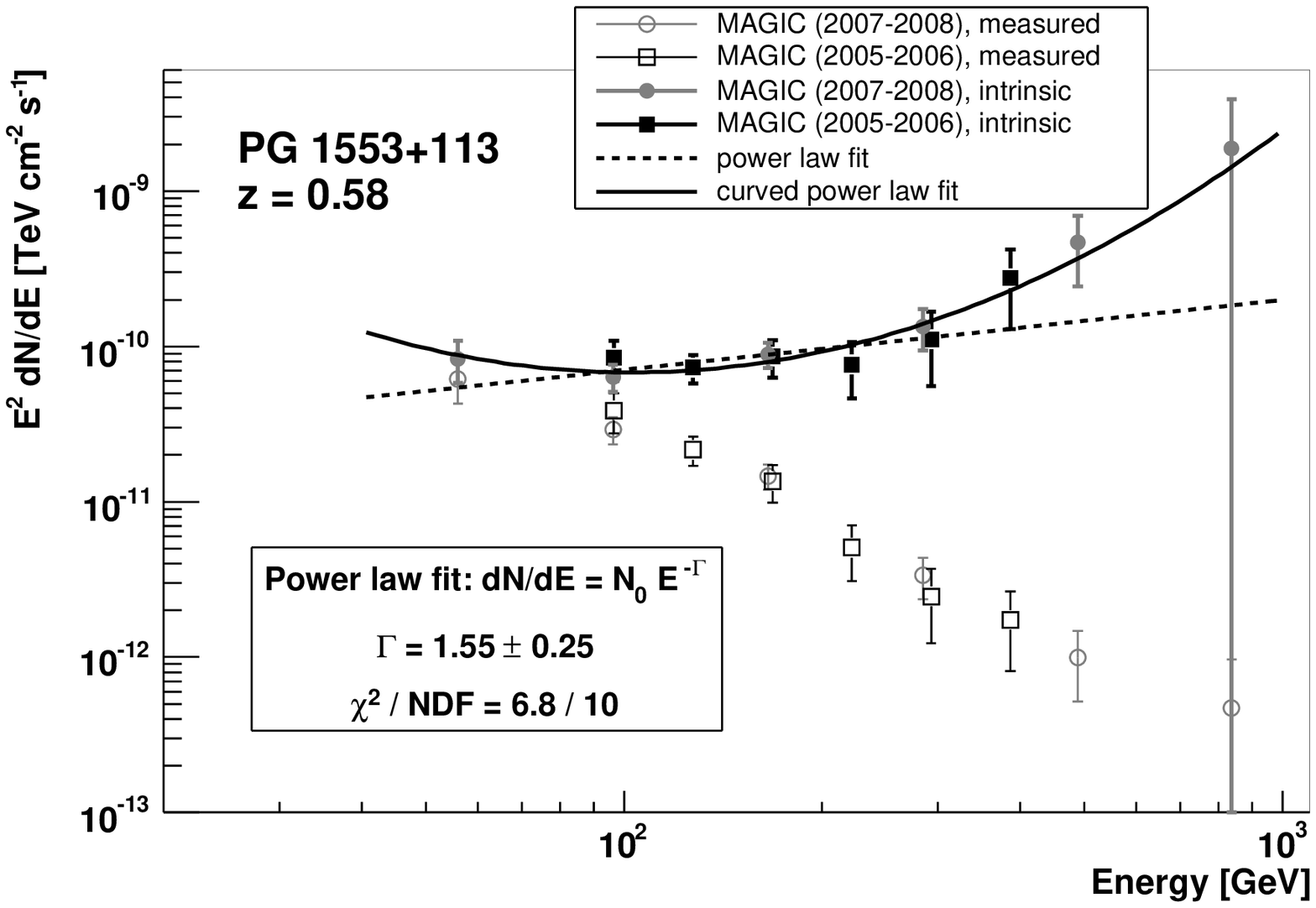} \label{2005_2008_deabsorbed}}
   }
   \caption{Constraints on the reshift of PG~1553+113.}
\end{figure*}

\subsection{Maximum intrinsic photon index}
We make the requirement that the power law index,
 $\Gamma_{int}$, obtained by 
fitting the deabsorbed spectrum with the simple power law of
 Eq.~\ref{simpeq}, plus twice its statistical error, is not harder
than $\Gamma_{int}^{st}=1.5$ in the standard scenario
and $\Gamma_{int}^{ex}=2/3$ in the extreme scenario case. 
The corresponding confidence level is 95$\%$.
We examined a wide range of redshift values z between
0.1 and 0.9 in steps of 0.02. Each time,
the intrinsic spectrum was determined 
using the model \cite{kneiske2004} and fitted.

The results and the corresponding 2$\sigma$ confidence
belt  are shown in Fig. \ref{scan}, thick
 blue line. The redshift limits obtained with this 
method are z$<$0.67 in the standard scenario case, and
z$<$0.80 in the extreme one.
The same analysis, performed by adopting the recent
EBL model \cite{franceschini2008},
 led to comparable values.

\subsection{Absence of a break in the intrinsic spectrum}
In order to test the absence of a break in the deabsorbed spectrum, 
we performed a likelihood ratio test between the hypothesis A, 
data fitted with the simple power law of Eq.\ref{simpeq},
 and hypothesis
B, data fitted with a curved power law, 
of index $-\alpha+\beta \ln(E)$.
This corresponds to a parabolic 
law in a $\log(E^2dN/dE)$ vs $\log(E)$
representation. More details about this method
 can be found in \cite{mazin2007}.

The resulted probability of the likelihood ratio test
is shown in Fig. \ref{2005_2008_deabsorbed}, thick black line.
With a confidence level of 95$\%$, the deabsorbed
 spectrum shows a break at redshift z$=$0.58.
The deep of the likelihood value at z$=$0.28 
indicates that at this redshift the intrinsic 
combined spectrum is a strict power law.
The requirement that the intrinsic spectrum
from PG~1553+113 does not show a pile up at high
energies, leads to an upper limit of z$<$0.58 on
its distance.
The same analysis performed by adopting the recent
EBL model, gives the limit z$<$0.60.

\section{Conclusions \label{sec:conc} }
Follow-up observations of PG~1553+113 at VHE carried out with the MAGIC 
Telescope in 2007 and 2008 have shown that
the source was in a quite steady state, within a factor 2.5,
 between 2005 and 2008 at these wavelengths. 
From the combined spectrum we have derived
an upper limit on the source redshift,
taking into account the absorption
of VHE photons with the EBL.

With a low EBL model, we showed 
that the intrinsic photon index $\Gamma_{int}$ becomes harder
than 1.5 at z$=$0.67. This can be considered as a 
robust upper limit on the redshift of PG~1553+113, 
taking into account the standard scenario of 
shock-accelerated electrons. Moreover, the
the spectral index of PG~1553+113 was recently 
 measured in the energy range 0.2-100 GeV
 by the Fermi LAT \cite{fermi2009} and has a value of
 $\Gamma=1.70\pm0.06$.

In case of extreme emission scenario, $\Gamma_{int}^{ex}=2/3$,
our limit becomes z$=0.80$. 
A break in the intrinsic spectrum
becomes evident at redshift z$=$0.58. The
pile up can either be interpreted as an upper limit
on the source redshift or as evidence for a second emission 
component in the VHE spectrum from PG~1553+113.

The limits obtained here are apparently higher than 
the ones
resulted in a previous work in which the same 
EBL model
was adopted \cite{mazin2007}. In that case,
however, data from different experiments were combined.
\newpage
The larger data sample used here, together with
the fact that the data were taken with a single  
experiment, leads to comparably smaller statistic 
and systematic errors in the flux evaluation, 
and results in a more realistic determination of the upper limits
on the source distance.

\section*{Acknowledgements\label{sec:conc} } 
 We would like to thank A.~Franceschini 
for providing the optical depth data, as well as
the Instituto de Astrofisica de 
Canarias for the excellent working conditions at the 
Observatorio del Roque de los Muchachos on La Palma. 
The support of the German BMBF and MPG, the Italian INFN 
and Spanish MICINN is gratefully acknowledged. 
This work was also supported by ETH Research Grant 
TH 34/043, by the Polish MNiSzW Grant N N203 390834, 
and by the YIP of the Helmholtz Gemeinschaft.

\end{document}